\documentclass[a4paper,12pt]{article}
\usepackage{appendix}
\usepackage[english]{babel}
\usepackage{color}
\usepackage{amsmath,amssymb,cite}%
\newtheorem{theorem}{Theorem}
\newtheorem{lemma}{Lemma}
\usepackage[bookmarks,hypertexnames=false,debug]{hyperref}
\usepackage{bookmark}
\usepackage[large,nohug,heads=vee]{diagrams}
\diagramstyle[labelstyle=\scriptstyle]
\def\eq#1{\begin{equation}#1\end{equation}}
\def\eqa#1{\begin{equation}\begin{aligned}#1\end{aligned}\end{equation}}
\def\eqs#1{\begin{eqnarray}#1\end{eqnarray}}
\def\qed{\vrule height0.6em width0.3em depth0pt\medskip}

\def\teqref#1{(T\ref{#1})}

\def\eeqref#1{(E\ref{#1})}
\def\dref#1{\ref{#1}}
\def\seq#1{\begin{equation*}#1\end{equation*}}
\def\seqs#1{\begin{equation*}\begin{split}#1\end{split}\end{equation*}}
\def\dfrac#1#2{\frac{\partial #1}{\partial #2}}
\def\subeq#1#2{\begin{subequations}\label{#1}\renewcommand{\theequation}{\theparentequation\alph{equation}} \begin{align}#2\end{align}\end{subequations}}

\font\Sets=msbm10
\def\Z {\hbox{\Sets Z}}
\def\C {\hbox{\Sets C}}
\def\proof{\noindent {\bf Proof.\ }}
\numberwithin{equation}{section}
\newcounter{transf}
\newenvironment{transform}[1]{\refstepcounter{transf}$$ #1\eqno{(\hbox{T}\thetransf)}$$} 

\newcounter{meq}
\newenvironment{mequ}[1]{\refstepcounter{meq}$$ #1\eqno{(\hbox{E}\themeq)}$$} 

\newcounter{dia}
\newtheorem{thm}{Theorem}[section]

\newtheorem{defn}{Definition}[section]

\title{\bf Classification of five-point differential-difference equations}
\author{{\bf R.N. Garifullin$^{1}$, R.I. Yamilov$^1$ and D. Levi$^2$}
\\$^1$ Institute of Mathematics, Ufa Scientific Center,\\ Russian Academy of Sciences,\\ 112 Chernyshevsky Street, Ufa 450008, Russian Federation
\\$^2$Department of Mathematics and Physics, Roma Tre University \\
and Sezione INFN  {\it Roma Tre},\\
Via della Vasca Navale 84, 00146 Rome, Italy\\
{\sl E-mails: rustem@matem.anrb.ru, RvlYamilov@matem.anrb.ru,}
\\ {\sl decio.levi@roma3.infn.it}}

\begin{document}
\maketitle
\abstract{Using the generalized symmetry method, we  carry out, up to autonomous point transformations,  the classification of integrable equations of a subclass of the autonomous five-point differential-difference equations. This subclass includes such well-known examples as the Itoh-Narita-Bogoyavlensky and the discrete Sawada-Kotera equations. The resulting list  contains 17 equations some of which seem to be new. We have  found non-point transformations relating most of the resulting equations among themselves and their generalized symmetries.}

\section{ Introduction}
\indent The generalized symmetry method uses the existence of generalized symmetries as an integrability criterion and allows one to classify integrable equations of a certain class. Using this method, the classification problem has been solved for some important classes of Partial Differential Equations  (PDEs) \cite{msy87,mss91}, of Differential-Difference Equations (D$\Delta$Es) \cite{asy00,y06}, and of Partial Difference Equations (P$\Delta$Es) \cite{ly11,gy12}. 

This is not the only integrability criterion introduced to produce integrable P$\Delta$Es. Using the Compatibility Around the Cube (CAC) technique introduced in \cite{n,bs02,n02}, Adler, Bobenko and Suris (ABS) \cite{ABS} obtained  a class of integrable equations on a quad graph. More recent results on this line of research can be found in \cite{abs2,boll,gsl16}. All equations obtained by ABS and their extensions have generalized symmetries which are integrable D$\Delta$Es, belonging  to the classification presented in \cite{y06,ly97} and given, in general, by D$\Delta$Es defined on  three-point lattices \cite{lpsy,gsl,gslqv}. 

Recently one can find many results  in which P$\Delta$Es defined on the square  but not determined in the ABS classification or in its extensions have generalized symmetries defined on more than three-point lattices \cite{a11,gy12,shl,mx13}. An extension of the  classification of the integrable P$\Delta$Es defined on a square is very difficult to perform. An alternative that seems more easy to perform is to  classify integrable five-point D$\Delta$Es
\eq{\label{y0}  \dot u_n=\Psi(u_{n+2},u_{n+1}, u_n, u_{n-1}, u_{n-2}).}
Here $\dot u_n$ is derivative of $u_n$ with respect to a continuous variable $t$.
Few results in this line of research are already known, see e.g. \cite{a14,a16,a16_1,gyl16}. The integrable P$\Delta$Es are then obtained as B\"acklund transformations of these D$\Delta$Es  \cite{lb,l81,gy15,ly09}.
The best known  integrable example in this class is the Ito-Narita-Bogoyavlensky (INB) equation \cite{i75,na82,bo88}: 
\eq{\label{INB0}\dot u_n=u_n(u_{n+2}+u_{n+1}-u_{n-1}-u_{n-2}).}

Volterra type equations 
\eq{\label{y1} \dot u_n=\Phi(u_{n+1}, u_n, u_{n-1})}
have been completely classified  \cite{y83}, and the resulting list of equations is quite big, see the details in the review article \cite{y06}. The classification of  five-point lattice equations of the form \eqref{y0} will contain equations coming from the classification of Volterra equations \eqref{y1}. 
For example, they appear if  we consider  equations of the form 
\eq{\label{y2} \dot u_n=\Phi(u_{n+2}, u_n, u_{n-2}).}
It is clear that if $u_n$ is a solution of \eqref{y2}, then the functions $\tilde u_k= u_{2k}$ and $\hat u_k=u_{2k+1}$ satisfy \eqref{y1} with $k$ instead of $n$. Eq. \eqref{y2} is just a three-point lattice equation equivalent to \eqref{y1}.
A second case is when we consider  generalized symmetries of \eqref{y1}. Any integrable Volterra type equation has a five-point  symmetry of the form \eqref{y0}. See the explicit results for Volterra type equations presented for example in \cite{y06,hlrw,sy88,sy90}. 

To avoid those two cases, which are included in  the classification of Volterra type equations and to simplify the problem, we consider here equations of the form
\eqa
{ \dot u_n = A(u_{n+1}, u_n,& u_{n-1}) u_{n+2}+B(u_{n+1}, u_n, u_{n-1}) u_{n-2}\\&+ C(u_{n+1}, u_n, u_{n-1}),\label{e1}}
where the form of $A$ and $B$ will be defined later, see \eqref{e2}.

Few equations of the Volterra classification \eqref{y1} are also included in the five-point classification \eqref{e1}. They are those equations which are 
linearly dependent on $u_{n+1}$ and $u_{n-1}$.  All of them are polynomial \cite{y06}. Such equations, rewritten in the form \eqref{y2}, belong to the class \eqref{e1}. Also their five-point symmetries are of the form \eqref{e1}. Moreover the majority of the examples  of D$\Delta$Es of the form \eqref{y0} known up to now   belong to the class \eqref{e1} \cite{a16_1,th96,mx13,gy12,bo88,b91,i75,na82,ap14}. So the class \eqref{e1} is not void.

The theory of the generalized symmetry method is well-developed in case of Volterra type equations \cite{y06} and it has been modified for the case of equations depending on 5 and more lattice points in \cite{a14,a16}. The classification problem of the class \eqref{e1} seems to be technically quite complicate. For this reason we use a simpler version of the method compared with the one presented in \cite{a14,a16}.

For equations analogous to  \eqref{INB0}, which are the first members of their hierarchies, the simplest generalized symmetry has the form \cite{ztof91,mx13,a11,gy12}:
\eq{u_{n,\tau}=G(u_{n+4},u_{n+3},u_{n+2},u_{n+1},u_{n},u_{n-1},u_{n-2},u_{n-3},u_{n-4}),\label{gen_sym}}
where $u_{n,\tau}$ denotes $\tau$-derivative of $u_n$. We will use the existence of such symmetry as an integrability criterion. 
 
The problem naturally splits into cases depending on the form of the functions $A$ and $B$. 
In this article we study the case when the autonomous D$\Delta$Es \eqref{e1} is such that $A$ and $B$ satisfy to the following conditions:
\begin{equation}\label{e2}
A \ne \alpha(u_{n+1},u_n) \alpha(u_n,u_{n-1}) , \quad B \ne\beta(u_{n+1},u_n) \beta(u_n,u_{n-1}) 
\end{equation} for any functions $\alpha$ and $\beta$ of their arguments.
We call this class the {\it Class~I}. Class~I includes such well-known examples as the
INB equation and the discrete Sawada-Kotera equation, see \cite{th96} and \eeqref{seva} in Section \ref{sec_list}.
The following simple criterion for checking conditions \eqref{e2} takes place:
\begin{equation}\label{e3}
\frac{\partial}{\partial u_n} \frac{a_{n+1} a_{n-1}}{a_n} \neq 0 , \qquad 
\frac{\partial}{\partial u_n} \frac{b_{n+1} b_{n-1}}{b_n} \neq 0 ,
\end{equation}where 
\seq{a_n=A(u_{n+1},u_n,u_{n-1}),\quad b_n=B(u_{n+1},u_n,u_{n-1}), 
 }  as we will show in  Theorem \ref{tt1}  in Section \ref{crit}. In the proof of this criterion an essential role is played by the fact that   \eqref{e1} is autonomous, i.e. has no explicit dependence on $n$.


In this article we present a complete list of equations of the Class~I possessing a generalized symmetry of the form \eqref{gen_sym}. Among them there are a few apparently new integrable examples. Then we show the  non-point transformations relating  most of resulting equations among themselves.

In Section \ref{sec_th} we discuss a theory of the generalized symmetry method suitable to solve our specific problem. In particular, in Section \ref{s_class} some integrability conditions are derived  and criteria for checking those conditions are proved in Section \ref{crit}. In Section \ref{sec_list} we present the obtained  list of integrable equations and the relations between those equations expressed in the form of the non-point transformations which are presented in Appendix A. In Section \ref{5} the generalized symmetries of the key equations of the resulting list are given. Section \ref{s_con} is devoted to some concluding remarks. 

\section{Theory}\label{sec_th}
To simplify the notation let us represent  \eqref{e1} as:
\eq{\dot u_n=a_nu_{n+2}+b_nu_{n-2}+c_n\equiv f_n,\label{ur_abc}} where \eqa{ a_n=A(u_{n+1},&u_n,u_{n-1}),\quad b_n=B(u_{n+1},u_n,u_{n-1}),\\ &c_n=C(u_{n+1},u_n,u_{n-1}).}
In \eqref{ur_abc} we require
 \eq{a_n\neq0,\quad b_n\neq0.\label{ogr_AB}} 
  For convenience we represent the symmetry  \eqref{gen_sym} as 
\eq{ u_{n,\tau}=g_n,\label{10s}}
with the restriction:
\eq{ \dfrac{g_n}{u_{n+4}} \ne 0, \qquad \dfrac{g_n}{u_{n-4}} \ne 0. \label{ogr_G}}

The compatibility condition for \eqref{ur_abc} and \eqref{10s} is
\eqa{&u_{n,\tau,t}-u_{n,t,\tau}\equiv D_t g_{n}-D_{\tau}f_{n}=0.\label{con_cond}}
As  \eqref{ur_abc} and \eqref{10s}  are autonomous, we can consider them and their compatibility condition \eqref{con_cond} at the point $n=0:$
\eq{D_t g_{0}=D_{\tau}f_0.\label{con_sh}} 
Here $D_t$ and $D_{\tau}$ are the operators of total differentiation with respect to $t$ and $\tau$  given respectively by:
\eq{D_t=\sum_{k\in\Z} f_k\dfrac{}{u_k}, \quad D_{\tau}=\sum_{k\in\Z} g_k\dfrac{}{u_k}.\label{dt}}

We assume as independent variables the  functions \eq{u_0,u_1,u_{-1},u_2,u_{-2},u_{3},u_{-3}.\ldots\label{i_var}}  Thus \eqref{con_sh} must be satisfied identically for all values of the independant variables \eqref{i_var}. Eq. \eqref{con_sh} depends on variables $u_{-6},u_{-5},\ldots,u_5,u_6$ and it is an overdetermined equation for the unknown function $g_0$, with given $f_0$.
Using a standard technique of the generalized symmetry method \cite{y06}, we can calculate $g_0$ step by step, obtaining conditions for the function $f_0$. 

\subsection{General case}
The first steps for the calculation of $g_0$ can be carried out with no  restriction on the form of the equation \eqref{ur_abc} given by the function $f_0$.

In fact, differentiating \eqref{con_sh} with respect to $u_6$, we obtain:
\eq{a_4\dfrac{g_0}{u_4}=a_0\dfrac{g_2}{u_6}.\label{***}}
Introducing the shift operator $T$: $Th_n=h_{n+1}$, we can rewrite \eqref{***}  as:
\seq{(T^2-1)\left(\frac1{a_0a_2}\dfrac{g_0}{u_4}\right)=0.}
The kernel of the operator $T^2-1$ in the autonomous case consists just of  constants \cite{y06}. Then, up to a $\tau$-scaling in \eqref{10s}, we can write
\eq{\label{gp4}\dfrac{g_0}{u_4}=a_0a_2.}
By differentiating \eqref{con_sh} with respect to $u_5$ and taking into account \eqref{gp4}, we obtain:
\eq{a_3\dfrac{g_0}{u_3}-a_0\dfrac{g_2}{u_5}-a_1a_3\dfrac{f_0}{u_1}+a_0a_2\dfrac{f_4}{u_5}=0.\label{du5}}
If we define  
\eq{h^+_0 = \dfrac{g_0}{u_3} - a_1\dfrac{f_0}{u_1}-a_0\dfrac{f_2}{u_3},\label{o_h}}
 then \eqref{du5} is equivalent to
\eq{a_3h^+_0=a_0h^+_2,\label{urh}}
where $h^+_2=T^2h^+_0$.

Then we can state the following Lemma:
\begin{lemma}If $h^+_0\ne0$, then there exists $\hat\alpha_n=\alpha(u_n,u_{n-1})$, such that ${a_0=\hat\alpha_1\hat\alpha_0}$, i.e. the equation is not of Class~I.\label{th_h}
\end{lemma}
\proof When we multiply \eqref{urh} by $\frac{h^+_1}{a_0a_1a_2a_3}$ we obtain
\seq{(T-1)\frac{h^+_0h^+_1}{a_0a_1a_2}=0.} 
As the kernel of the operator $T-1$ consists of constants, we have \eq{\frac{h^+_0h^+_1}{a_0a_1a_2}=\eta^2\neq0.\label{eta}} Eq. \eqref{eta} is equivalent to
$a_0=\hat\alpha_1\hat\alpha_0,$ where $\hat\alpha_0=\frac{h^+_0}{\eta a_1}.$ \qed

There are two possibilities:
\begin{itemize}
\item Case 1. $a_0\neq\hat\alpha_1\hat\alpha_0$ for any $\hat\alpha_n=\alpha(u_n,u_{n-1})$, cf. \eqref{e3}. Then $h^+_0=0$ due to Lemma \ref{th_h}.
\item Case 2. $a_0=\hat\alpha_0\hat\alpha_1$ for some $\hat\alpha_n=\alpha(u_n,u_{n-1})$. Then we can find from \eqref{urh} that $h^+_0=\mu^+\hat\alpha_0\hat\alpha_1\hat\alpha_2$ with a constant $\mu^+$.
\end{itemize}
\noindent In both cases \eqref{o_h} gives us $\dfrac{g_0}{u_3}$.

In quite similar way, differentiating $\eqref{con_sh}$ with respect to $u_{-6}$ and $u_{-5}$, we get a set of relations analogous to \eqref{gp4} and \eqref{urh}. Namely,
\eq{\label{gm4}\dfrac{g_0}{u_{-4}}=\nu b_0b_{-2},}
\eq{b_{-3} h^-_0 =b_0h^-_{-2},\label{urhh}} where $\nu\ne0$ is a constant, and
\eq{h^-_0 = \dfrac{g_0}{u_{-3}} - \nu b_{-1}\dfrac{f_0}{u_{-1}}-\nu b_0\dfrac{f_{-2}}{u_{-3}}.\label{o_hh}}

As a consequence of a Lemma similar to Lemma \ref{th_h} we get  two cases:
\begin{enumerate}
\item $b_0\neq\hat\beta_0\hat\beta_{-1}$ for any $\hat\beta_n=\beta(u_{n+1},u_{n})$, then $h^-_0=0$.
\item  $b_0=\hat\beta_0\hat\beta_{-1}$, then we can find from \eqref{urhh} that $h^-_0=\mu^-\hat\beta_0\hat\beta_{-1}\hat\beta_{-2}$ with a constant $\mu^-$.
\end{enumerate}
\noindent In both cases \eqref{o_hh} provides us $\dfrac{g_0}{u_{-3}}$.

So the results presented in this subsection provide a natural frame for splitting further calculation of $g_0$  into several different cases. In the following  in this paper we consider Class~I in which condition \eqref{e2} is satisfied and therefore $h_0^{\pm}=0$. 

\subsection{Class~I}\label{s_class}
When condition \eqref{e2} is satisfied and therefore $h^+_0=h^-_0=0$, then due to (\ref{o_h},\ref{o_hh})
\eq{\dfrac{g_0}{u_3} = a_1\dfrac{f_0}{u_1}+a_0\dfrac{f_2}{u_3}\label{gp3},}
\eq{\dfrac{g_0}{u_{-3}} = \nu b_{-1}\dfrac{f_0}{u_{-1}}+\nu b_0\dfrac{f_{-2}}{u_{-3}}\label{gm3}.} 
Partial derivatives $\dfrac{g_0}{u_4},\dfrac{g_0}{u_{-4}} $ are always given by (\ref{gp4}, \ref{gm4}).

Differentiating \eqref{con_sh} with respect to $u_4$ and $u_{-4}$ and introducing the functions:
\eq{q^+_0=\frac1{a_0}\dfrac{g_0}{u_2}-D_t \log a_0-\dfrac{f_0}{u_0}-\dfrac{f_2}{u_2}-\frac1{a_0}\dfrac{f_0}{u_1}\dfrac{f_1}{u_2},\label{qp}}
\eq{q^-_0=\frac1{\nu b_0}\dfrac{g_0}{u_{-2}}-D_t \log b_0-\dfrac{f_0}{u_0}-\dfrac{f_{-2}}{u_{-2}}-\frac1{b_0}\dfrac{f_0}{u_{-1}}\dfrac{f_{-1}}{u_{-2}},\label{qm}}
 we obtain, up to a common factor $a_0\,a_2$, the relation:
\eq{2D_t \log a_0=q^+_2-q^+_0,\label{la}} 
and up to a common factor $\nu b_0\,b_{-2}$, the relation:
\eq{2D_t \log b_0=q^-_{-2}-q^-_0.\label{lb}}
It is evident that
\eq{q^+_0=q^+_0(u_1,u_0,u_{-1},u_{-2},u_{-3}),\quad q^-_0=q^-_0(u_3,u_2,u_{1},u_{0},u_{-1}).\label{qpm}} 
The relations \eqref{la} and \eqref{lb} have the form of conservation laws and are necessary conditions for the integrability. If  a symmetry \eqref{10s} exists for \eqref{ur_abc}, then there must exist some functions $q_n^+,q_n^-$ of the form \eqref{qpm} such that relations \eqref{la} and \eqref{lb} are satisfied. 

The integrability conditions are formulated in terms of equation \eqref{ur_abc} only, i.e.  for an integrable equation \eqref{ur_abc}, there must exist functions $q^\pm_n$ of the form \eqref{qpm} satisfying the relations \eqref{la} and \eqref{lb}.

If, for a given equation \eqref{ur_abc},  conditions \eqref{la} and \eqref{lb} are satisfied and the functions $q_n^\pm$ are known, then partial derivatives $\dfrac{g_0}{u_2},\dfrac{g_0}{u_{-2}}$ can be found from (\ref{qp}, \ref{qm}). In this case the right hand side of symmetry \eqref{10s} is defined up to one unknown function of 3 variables:
\eq{\label{fpsi}\psi(u_{n+1},u_n,u_{n-1}).} This function can be found directly from the compatibility condition \eqref{con_sh}.

In this way we can carry out the classification of the equations of Class~I. On the first stage we use the integrability conditions (\ref{la}, \ref{lb}). Then we define the symmetry up to function \eqref{fpsi} and try to find it from the compatibility condition.

\subsection{Criteria for checking the integrability conditions  (\ref{la}, \ref{lb})} \label{crit}
Let us now  explain how to use the integrability conditions (\ref{la}, \ref{lb}).  More precisely, we present some criteria for checking those conditions. We will also prove the criterion \eqref{e3}.

Let us  introduce for any function
\eq{\label{varphi}\varphi=\varphi(u_{m_1},u_{m_1-1},\ldots,u_{m_2}),\quad m_1\geq m_2,} 
 the formal variational derivative:
\eq{\frac{\delta \varphi}{\delta u_0}=\sum_{k=m_2}^{m_1}T^{-k}\dfrac{\varphi}{u_k}=\dfrac{}{u_0}\sum_{k=m_2}^{m_1}T^{-k}\varphi,\label{v1}} see e.g. \cite{y06}, as well as its adjoint version: 
\eq{\frac{\bar\delta \varphi}{\bar\delta u_0}=\sum_{k=m_2}^{m_1}(-1)^kT^{-k}\dfrac{\varphi}{u_k}=\dfrac{}{u_0}\sum_{k=m_2}^{m_1}(-1)^kT^{-k}\varphi,\label{v2}} 
Then we can state the following Lemma: 
\begin{lemma}\label{l_c} The following statements are true: \eqs{\qquad \frac{\delta \varphi}{\delta u_0}=0\quad &\mathrm{iff}& \quad\varphi=\kappa+(T-1)\omega\label{f_d},
\\ \qquad\frac{\bar\delta \varphi}{\bar\delta u_0}=0\quad &\mathrm{iff}& \quad\varphi=(T+1)\omega,\label{ad_f_d}} where $\kappa$ is a constant, while $\omega$ is a function of a finite number of independent variables \eqref{i_var}.
\end{lemma}
This Lemma implies  that if the function $\varphi$ is such that its variational derivative with respect to $u_0$ is zero, then it can be represented in terms of some constant $\kappa$ and  a function $\omega$ depending on a finite number of independent variables. 

\medskip

\proof The proof of \eqref{f_d} and \eqref{ad_f_d} are similar. The proof of \eqref{f_d} is given in \cite{y06}. So  we present here only the proof of \eqref{ad_f_d}.

If function $\varphi$ can be expressed as $(T+1)\omega$ and $m_1>m_2$, then $\omega=\omega(u_{m_1-1},\ldots,u_{m_2})$. For such functions $\varphi$ we have: 
\seqs{\frac{\bar\delta \varphi}{\bar\delta u_0}&=\dfrac{}{u_0}\sum_{k=m_2}^{m_1}(-1)^kT^{-k}(T+1)\omega\\&=\dfrac{}{u_0}((-1)^{m_2}T^{-m_2+1}\omega+(-1)^{m_1}T^{-m_1}\omega)=0,} where the last equality is identically satisfied,  as both terms in the last sum do not depend on $u_0$.

In the case when $m_1=m_2$ the assumption  $\varphi=(T+1)\omega$ implies that the function $\omega$ and therefore $\varphi$ are constant functions, and hence $\frac{\bar\delta \varphi}{\bar\delta u_0}=0.$

Let us now assume that   $\varphi$ satisfies the equation $\frac{\bar\delta \varphi}{\bar\delta u_0}=0$ with $m_1>m_2$. Then it follows that the following equation is also satisfied
$$\dfrac{}{u_{m_1}}(-1)^{m_2}T^{m_2}\frac{\bar\delta \varphi}{\bar\delta u_0}=0.$$ Explicitating the variational derivative  we have 
$$\dfrac{^2\varphi}{u_{m_1}\partial u_{m_2}}=0.$$ So we can represent $\varphi$ as 
\seqs{\varphi=\theta(u_{m_1},u_{m_1-1},\ldots,u_{m_2+1})+\rho(u_{m_1-1},u_{m_1-2},\ldots,u_{m_2})\\
=(T+1)\rho+\hat\varphi(u_{m_1},u_{m_1-1},\ldots,u_{m_2+1}),}  with $
\hat\varphi=\theta-T\rho.$

As $\frac{\bar\delta \varphi}{\bar\delta u_0}=\frac{\bar\delta \hat\varphi}{\bar\delta u_0}=0$, we have reduced the problem to a lower number of variables. 
Repeating this procedure, we finally obtain $$\varphi=\tilde\varphi(u_{m_1})+(T+1)\tilde \omega.$$  Then it is easy to see that $$\dfrac{\tilde\varphi}{u_{m_1}}=(-1)^{m_1}T^{m_1}\frac{\bar\delta\tilde \varphi}{\bar\delta u_0}=0,$$ i.e. $\tilde \varphi$ is a constant, and $\varphi=(T+1)(\tilde\omega+\tilde\varphi/2).$ \qed  
\medskip 

This Lemma allows us to formulate and prove the following theorems:
\begin{theorem}  The request that conditions \eqref{e2} are satisfied is equivalent to \eqref{e3}. \label{tt1}\end{theorem}
\proof The condition $A \ne \alpha_n \alpha_{n+1}$  can be rewritten in terms of the function $\log a_n=\log A$ as $$\log A\not =(T+1)\log\alpha(u_n,u_{n-1})$$ 
or, in equivalent form,
$$\frac{\bar\delta }{\bar\delta u_0}\log a_n=-\dfrac{}{u_n}\log\frac{a_{n+1}a_{n-1}}{a_n}\neq0.$$ Their equivalence follows from Lemma \ref{l_c}. In an analogous way we get the result for $b_n$. \qed

\begin{theorem} \label{tt2} The conditions  \eq{\label{ca}\frac{\delta }{\delta u_0}D_t\log a_0=0,\quad\frac{\bar\delta }{\bar\delta u_0}D_t\log a_0=0}
imply \eq{2D_t \log a_0=\kappa^++(T^2-1)q^+_0,\quad \kappa^+\in\C,\label{lla}}  and vice versa. \end{theorem}
\proof
As \eqref{lla} can be rewritten in the form
\seq{2D_t \log a_0=\kappa^++(T-1)(T+1)q^+_0=(T+1)(\kappa^+/2+(T-1)q^+_0),} we see that \eqref{lla} implies \eqref{ca} due to Lemma \ref{l_c}. 

Let us start from \eqref{ca}. From the first relation of \eqref{ca} and statement \eqref{f_d}, we obtain \seq{2D_t \log a_0=\kappa^++(T-1)Q_0^+=\kappa^++(T+1)Q_0^+-2Q_0^+,} therefore ${\frac{\bar\delta }{\bar\delta u_0}Q_0^+=0}$ due to the second part of \eqref{ca}. Writing  $Q_0^+=(T+1)q_0^+$, we get  \eqref{lla}. \qed

\begin{theorem} \label{tt3} The conditions  \eq{\frac{\delta }{\delta u_0}D_t\log b_0=0,\quad\frac{\bar\delta }{\bar\delta u_0}D_t\log b_0=0\label{cb}}
imply \eq{2D_t \log b_0=\kappa^-+(T^{-2}-1)q^-_0,\quad \kappa^-\in\C,\label{llb}} and vice versa.
\end{theorem}
As $(T^{-2}-1)q_0^-=(T^2-1)(-q_{-2}^-)$, we see that conditions \eqref{lla} and \eqref{llb} are of the same type and so  the proof of Therem \ref{tt3} repeats that of  Theorem \ref{t2}.

Let us simplify the integrability conditions \eqref{ca} and \eqref{cb}.
To do so we use a new representation for the variational derivatives \eqref{v1} and \eqref{v2}:
\eq{\label{var3}\frac{\delta \varphi}{\delta u_0}=\sum_{j=-m_1}^{-m_2}\dfrac{}{u_0}T^j\varphi, \quad \frac{\bar\delta \varphi}{\bar\delta u_0}=\sum_{j=-m_1}^{-m_2}(-1)^j\dfrac{}{u_0}T^j\varphi.}
Moreover in place of  conditions \eqref{ca} and \eqref{cb} we use their linear combinations:
\seq{\frac12\left(\frac{\delta }{\delta u_0}+\frac{\bar\delta }{\bar\delta u_0}\right)D_t\log a_0=0,\quad\frac12\left(\frac{\delta }{\delta u_0}-\frac{\bar\delta }{\bar\delta u_0}\right)D_t\log a_0=0,}
\seq{\frac12\left(\frac{\delta }{\delta u_0}+\frac{\bar\delta }{\bar\delta u_0}\right)D_t\log b_0=0,\quad\frac12\left(\frac{\delta }{\delta u_0}-\frac{\bar\delta }{\bar\delta u_0}\right)D_t\log b_0=0.}
As $$D_t \log a_0 = \frac{f_1}{a_0} \frac{\partial a_0}{\partial u_1} +\frac{f_0}{a_0} \frac{\partial a_0}{\partial u_0}+\frac{f_{-1}}{a_0} \frac{\partial a_0}{\partial u_{-1}},$$ then we use \eqref{var3} with $m_1=-m_2=3$  and we see that the conditions \eqref{ca} are equivalent to 
\subeq{dla}{\dfrac{}{u_0}D_t\log (a_{-2}a_0a_{2})=0,\label{dla1}\\ \dfrac{}{u_0}D_t\log (a_{-3}a_{-1}a_1a_{3})=0.\label{dla2}}
Similarly, the conditions \eqref{cb} are equivalent to 
\subeq{dlb}{\dfrac{}{u_0}D_t\log (b_{-2}b_0b_{2})=0,\label{dlb1} \\ \dfrac{}{u_0}D_t\log (b_{-3}b_{-1}b_1b_{3})=0.\label{dlb2}}

These are explicit and simple conditions for checking the integrability conditions \eqref{la} and \eqref{lb}. When these conditions are satisfied for a given equation \eqref{ur_abc}, one has to write down the representations \eqref{lla} and \eqref{llb} and check that they are satisfied, i.e. $\kappa^\pm=0$.

\subsection{The beginning of classification}\label{th_res}

We have explicit formulae (\ref{gp4}, \ref{gm4}, \ref{gp3}, \ref{gm3}) for the partial derivatives \eq{\label{diff_g}\dfrac{g_0}{u_4},\quad \dfrac{g_0}{u_{-4}},\quad \dfrac{g_0}{u_3},\quad \dfrac{g_0}{u_{-3}}} and implicit definitions for \eq{\label{dg2}\dfrac{g_0}{u_2},\ \dfrac{g_0}{u_{-2}}} given by the relations (\ref{qp}, \ref{qm}, \ref{la}, \ref{lb}).
The compatibility of the partial derivatives \eqref{diff_g}  gives no conditions. The compatibility of \eqref{dg2}  with each other  and with \eqref{diff_g} provides three restrictions. 
In fact, differentiating, for example, \eqref{la} with respect to $u_{-3}$ and using \eqref{qp}, we get $\dfrac{^2g_0}{u_{-3}\partial u_{2}}$ in explicit form. This has to be equal to the differentiation with respect to $u_2$  of $\dfrac{g_0}{u_{-3}}$ given by \eqref{gm3}. 
In this way, using the compatibility of $\dfrac{g_0}{u_{-3}}$ and $\dfrac{g_0}{u_{2}}$, $\dfrac{g_0}{u_{3}}$ and $\dfrac{g_0}{u_{-2}}$, $\dfrac{g_0}{u_{-2}}$ and $\dfrac{g_0}{u_{2}}$, we are led to
\eqs{(\nu+1)\dfrac{b_0}{u_1}=0,\quad (\nu+1)\dfrac{a_0}{u_{-1}}=0,\quad (\nu+1)\dfrac{}{u_{0}}\frac{a_0}{b_0}=0,\label{c_nu}}
where $a_0,\,b_0$ and $\nu$ are  defined in \eqref{ur_abc} and by \eqref{gm4}. 

Conditions \eqref{c_nu} allow one to split the classification problem into  two cases: 
\begin{enumerate}
\item The case $\nu\neq -1$ which is simpler and leads to  \eeqref{Bur}.
\item The case when $\nu=-1$.
\end{enumerate}
 
Using the explicit integrability conditions (\ref{dla}) and \eqref{dlb} for $\nu=-1$, we can get two results given in the following Theorems \ref{tt4} and \ref{tt5} which are formulated in terms of equation \eqref{ur_abc} only.

\begin{theorem} \label{tt4}
If an equation \eqref{ur_abc} satisfies conditions \eqref{dla} and \eqref{dlb}, then its coefficients must have the form:
\eq{a_0=a^{(1)}(u_1,u_0)a^{(2)}(u_0,u_{-1}),\quad b_0=b^{(1)}(u_1,u_0)b^{(2)}(u_0,u_{-1}).\label{ab}}
\end{theorem}
\proof This Theorem is divided in two parts. One related to $a_0$ and the second one to $b_0$. We will prove in the following the first part only as the second can be derived in the same way.

The first relation of \eqref{ab} is equivalent to require that  
\eq{a^{(3)}=\dfrac{^2}{u_{-1}\partial u_1}\log a_0\equiv 0.\label{a3}} 
Let us define $a^{(4)}$ and $a^{(5)}$ to be the functions on the left hand side of \eqref{dla1} and \eqref{dla2}.
 At first we see, that 
\seqs{\dfrac{a^{(3)}}{u_1}=\frac{1}{b_{-1}}T^{-3}\dfrac{^2}{u_4^2}a^{(5)}=0,\quad
\dfrac{a^{(3)}}{u_{-1}}=\frac{1}{a_{1}}T^{-1}\dfrac{^2}{u_4\partial u_0}a^{(5)}=0,} i.e. $a^{(3)}=a^{(3)}(u_0).$ 

Let us suppose, reductio ad absurdum, that $a^{(3)}\ne0.$ As a consequence we have
\seq{\dfrac{^2a_0}{u_{-1}^2}=\frac{1}{2Ta^{(3)}}T^{-1}\dfrac{^3 a^{(4)}}{u_3^2\partial u_0}=0,}
and this implies the identity \seq{\dfrac{}{u_{-1}}\left(a_0^2 a^{(3)}\right)=0.\label{a33}}  As $a^{(3)}=a^{(3)}(u_0)\neq 0$ by assumption, we find  that $\dfrac{a_0}{u_{-1}}=0$. Now, due to the definition of $a^{(3)}$ given in \eqref{a3}, we get $a^{(3)}=0$. This gives a contradiction which proves the Theorem. \qed

Taking into account the results just obtained in Theorem \ref{tt4}, we can prove the following Theorem:
\begin{theorem} \label{tt5}
If an equation \eqref{ur_abc} belongs to the Class~I and satisfies conditions (\ref{dla}, \ref{dlb}) and \eqref{ab}, then for the functions $a^{(2)}$ and $b^{(1)}$ one has:
\eq{\dfrac{^2a^{(2)}}{u_{-1}^2}=0,\qquad \dfrac{^2b^{(1)}}{u_1^2}=0.\label{a2b1}}
\end{theorem}
\proof We prove just the first equality in \eqref{a2b1}, as the second one can be proved in exactly the same way.

By direct calculation we derive from \eqref{dla2} and \eqref{ab} that
\seq{\dfrac{^2a^{(2)}}{u_{-1}^2} \cdot \dfrac{^2 \log a^{(1)}}{u_0 u_{1}}=\dfrac{}{u_1}\left(\frac1{a^{(1)}}T^{-1}\dfrac{^2a^{(5)}}{u_0\partial u_3}\right)=0.}
In this proof $a^{(5)}$ and $a^{(4)}$  are the same as in previous one.
So, either $\dfrac{^2a^{(2)}}{u_{-1}^2}=0$ or $\dfrac{^2 \log a^{(1)}}{u_0 u_{1}}=0$. In the first instance the Theorem has been proved. 

Let us then assume that $\dfrac{^2a^{(2)}}{u_{-1}^2} \ne 0$. Changing notations for the functions $a^{(1)}$ and $a^{(2)}$, we get $a^{(1)}=a^{(1)}(u_{1})$. By a straightforward calculation we can show that $$\dfrac{( a^{(2)} T^{-2} a^{(1)})}{u_{-1}}=\frac{a^{(2)}T^{-2}a^{(1)} }{T^{-1}a^{(1)}}\dfrac{^2T^{-2}a^{(4)}}{u_{-2}\partial u_1}\Big/T^{-1}\dfrac{^2a^{(2)}}{u_{-1}^2}=0,$$ i.e. $a^{(2)}=\frac{\theta(u_0)}{a^{(1)}(u_{-1})}$. The function $\theta$ is constant, as $$\frac{d \theta}{d u_0}=\frac{\theta }{a^{(1)}}\dfrac{^2 T^{-1}a^{(5)}}{u_2\partial u_{-1}}\Big/\dfrac{^2a^{(2)}}{u_{-1}^2}=0.$$ The condition \eqref{e3} is not satisfied for $a_n=\theta a^{(1)}(u_{n+1})/a^{(1)}(u_{n-1})$, and the resulting equation would not be of Class~I.  This is in contradiction with one of the hypothesis of this Theorem. So this instance is not possible. 
\qed


\section{Complete list of integrable equations}\label{sec_list}
In this Section we present the complete list of integrable equations of Class~I together with the non-point relations between them. These equations are referred by special numbers \eeqref{Vol}-\eeqref{sroot}. Some of the obtained equations seem to be new. 

The classification is usually carried out in two steps: at first one finds all integrable equations of a certain class up to point transformations, then one searches for non-point transformations which link the different resulting equations. In this paper we use autonomous point transformations which, because of the specific form \eqref{e1} of the equations, are linear transformations with constant coefficients:\eq{\label{point_t}\hat u_0=c_1u_0+c_2,\quad  \hat t=c_3t,\qquad c_1c_3\neq0.}
Non-point transformations we use here are transformations of the form
\eq{\label{transf}\hat u_0=\varphi(u_k,u_{k-1},\ldots,u_m),\ \ k>m,} and their compositions. 

The equation \eqref{y0} is transformed into
\eq{\label{ty0}   \hat u_{0,t}=\hat \Psi(\hat u_{2},\hat u_{1}, \hat u_0, \hat u_{-1}, \hat u_{-2})}
by \eqref{transf},
if for any solution $u_n$ of \eqref{y0}, formula \eqref{transf} provides a solution $\hat u_n$ of \eqref{ty0}. To check this, we substitute \eqref{transf} into \eqref{ty0}, differentiate with respect to $t$ in virtue of equation \eqref{y0} and check that the resulting relation  is satisfied identically for all values of independent variables \eqref{i_var}.

We see that transformation \eqref{transf} is explicit in one direction. If an equation $A$ is transformed into $B$ by a transformation \eqref{transf}, then this transformation has the {\it direction} from $A$ to $B$ and we will write in diagrams below $A\to B$, so indicating  the direction where it is explicit.

The complete list has been obtained up to autonomous linear point transformations \eqref{point_t}. The precise result is formulated as follows:

\begin{theorem} If a nonlinear equation of the form (\ref{ur_abc} -- \ref{ogr_AB}) belongs to Class~I, given by restrictions \eqref{e3}, and has a generalized symmetry (\ref{gen_sym}, \ref{10s}, \ref{ogr_G}), then up to point transformation \eqref{point_t} it is equivalent to one of the equations \eeqref{Vol}-\eeqref{sroot}.  Any of the equations \eeqref{Vol}-\eeqref{sroot} has a generalized symmetry of the form (\ref{gen_sym}, \ref{10s}, \ref{ogr_G}).
\end{theorem}

For easier understanding of the results, we split the complete list into smaller Lists 1-5. In each List the equations are related to each other by non-point transformations which are discussed  in Appendix \ref{tran}. For each of these lists we show relations between equations in diagrams, where the transformations  \eqref{transf} are shown by arrows. Transformations used in those diagrams have special numbers \teqref{tr_sum}, \teqref{tr_raz}, \ldots and are listed in Appendix \ref{A4}. Generalized symmetries for key equations of the complete list are presented in the following Section.

\bigskip\bigskip
{ \centerline{{\bf List 1.} Equations related to the  Volterra equation}}
\medskip
\begin{mequ}{\dot u_0=u_0(u_{2}-u_{-2})\label{Vol} }\end{mequ}
\begin{mequ}{\dot u_0=u_0^2(u_{2}-u_{-2})\label{Vol0} }\end{mequ}
\begin{mequ}{\dot u_0=(u_0^2+u_0)(u_{2}-u_{-2})\label{Vol1} }\end{mequ}
\begin{mequ}{\dot u_0=(u_{2}+u_{1})(u_0+u_{-1})-(u_1+u_0)(u_{-1}+u_{-2})\label{Vol_mod} }\end{mequ}
\begin{mequ}{\dot u_0=(u_{2}-u_{1}+a)(u_0-u_{-1}+a)+(u_1-u_0+a)(u_{-1}-u_{-2}+a)+b\label{Vol_mod1} }\end{mequ}
\begin{mequ}{\dot u_0=u_{2}u_{1}u_0(u_0u_{-1}+1)-(u_1u_0+1)u_0u_{-1}u_{-2}+u_0^2(u_{-1}-u_{1})\label{Vol2} }\end{mequ}

\bigskip
\noindent All equations of List 1 are transformed into \eeqref{Vol} as  shown in Diagram \dref{dVol}. 
\eq
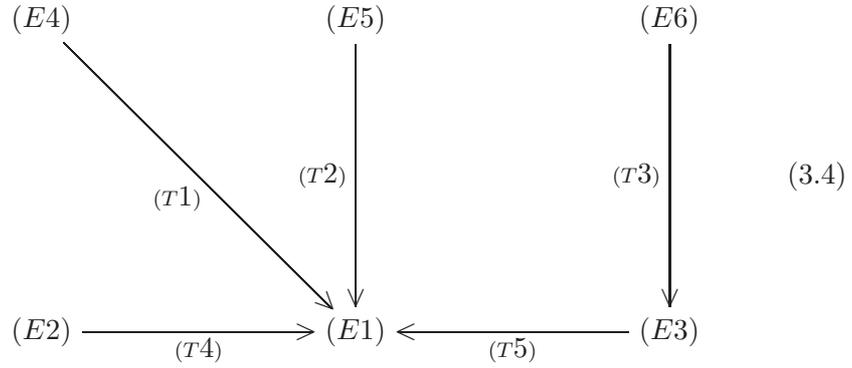
{\begin{diagram} \label{dVol} 
\eeqref{Vol_mod}& &\eeqref{Vol_mod1}& &\eeqref{Vol2}\\
&\rdTo{}{\teqref{tr_sum}}&\dTo{\teqref{tr_raz}}& &\dTo{\teqref{tr_1}}\\
\eeqref{Vol0}&\rTo{}{\teqref{tr_pr2}}&\eeqref{Vol}&\lTo{}{\teqref{tr_m2}}&\eeqref{Vol1}
\end{diagram}}

All transformations here are linearizable except for \teqref{tr_m2} which is of Miura type. The notions of linearizable and of Miura type transformations are discussed in Appendix \ref{A1}. Transformations $\tilde u_k=u_{2k}$ or $\tilde u_k=u_{2k+1}$ turn equations \eeqref{Vol}-\eeqref{Vol1} into the well-known Volterra equation and its modifications in their standard form. Transformations \teqref{tr_pr2} and \teqref{tr_m2} also turn into the standard ones, see e.g. \cite{y06}.

\bigskip\bigskip
{ \centerline{{\bf List 2.} Linearizable equations}}
\medskip
\begin{mequ}{\dot u_0=(T-a)\left(\frac{(u_1+au_0+b)(u_{-1}+au_{-2}+b)}{u_{0}+au_{-1}+b}+u_0+au_{-1}+b\right)+cu_0+d\label{Bur2} }\end{mequ}
\begin{mequ}{\dot u_0=\frac{u_2u_0}{u_1}+u_1-a^2\left(u_{-1}+\frac{u_0u_{-2}}{u_{-1}}\right)+cu_0\label{Bur} }\end{mequ}\noindent In both equations $ a\neq0,\ (a+1)d=bc$.

\bigskip
Both equations of List 2 are related to the linear one:
\begin{equation}\label{lin_eq} \dot u_0=u_2-a^2 u_{-2}+cu_0/2 \end{equation}
as it is shown in Diagram \dref{diag_lin}.
\eq{\begin{diagram} \label{diag_lin}
\eqref{lin_eq}&\rTo{\teqref{tr_1}}&\eeqref{Bur}&\lTo{\teqref{tr_summ}}&\eeqref{Bur2}
\end{diagram}}
However  \eeqref{Bur2} is related to the linear equation \eqref{lin_eq}  by a linearizable transformation which is implicit in both directions.

\bigskip\bigskip
{\centerline{{\bf List 3.} Equations of the relativistic Toda type}}
\medskip
\begin{mequ}{\dot u_0=(u_0-1)\left(\frac{u_2(u_1-1)u_0}{u_1}-\frac{u_0(u_{-1}-1)u_{-2}}{u_{-1}}-u_1+u_{-1}\right)\label{our1} }\end{mequ}
\begin{mequ}{\dot u_0=\frac{u_2u_1^2u_0^2(u_0u_{-1}+1)}{u_1u_0+1}-\frac{(u_1u_0+1)u_0^2u_{-1}^2u_{-2}}{u_0u_{-1}+1}\label{our2} }\end{mequ}
\vspace{-0.2cm}
$$\qquad\qquad-\frac{(u_1-u_{-1})(2u_1u_0u_{-1}+u_1+u_{-1})u_0^3}{(u_1u_0+1)(u_0u_{-1}+1)} $$
\bigskip
The equations of List 3 are related to the following one:
\begin{equation}{\dot u_0=(u_1u_0-1)(u_0u_{-1}-1)(u_2-u_{-2})\label{our} }\end{equation}
as it is shown in Diagram \dref{diag_our}.

\eq{\begin{diagram} 
\eqref{our}&\rTo{\teqref{tr_1}}&\eeqref{our1}&\lTo{\teqref{tr_prod_m}}&\eeqref{our2}\label{diag_our} 
\end{diagram}}
Eq. \eqref{our}, which is out of Class~I, is known and well-studied \cite{gy12,gmy14}. It is shown in \cite{gmy14} that it is of the relativistic Toda type. It is natural that equations \eeqref{our1} and \eeqref{our2}, of Class~I, being related to \eqref{our} are of the same type. Equation \eeqref{our1} obtained by transformation \teqref{tr_1} from \eqref{our} is also known, see \cite{a16_1}. The relation between \eeqref{our2} and \eqref{our} is linearizable, but completely implicit.

\bigskip\bigskip
{ \centerline{{\bf List 4.} Equations related to the INB}}
\medskip
\begin{mequ}{\label{INB}\dot u_0=u_0(u_2+u_1-u_{-1}-u_{-2})}\end{mequ}
\begin{mequ}{\label{mod_INB}\dot u_0=(u_{2}-u_{1}+a)(u_0-u_{-1}+a)+(u_1-u_0+a)(u_{-1}-u_{-2}+a)
\\+(u_1-u_0+a)(u_0-u_{-1}+a)+b }\end{mequ}
\begin{mequ}
{\dot u_0 = (u_0^2+au_0)(u_2u_1-u_{-1}u_{-2})\label{mikh1}}
\end{mequ}
\begin{mequ}{\label{eq1}\dot u_0 = (u_1-u_0)(u_0-u_{-1})\left(\frac {u_2}{u_1}-\frac{u_{-2}}{u_{-1}}\right)}\end{mequ}
\bigskip
\noindent All equations of List 4 are transformed into the INB equation \eeqref{INB} as it is shown in Diagram \dref{dINB}.
\eq{\begin{diagram} \label{dINB}
& &\eeqref{mikh1}& &\\
&&\dTo{\teqref{minb}}& &\\
\eeqref{mod_INB}&\rTo{\teqref{tr_raz}}&\eeqref{INB}&\lTo{\teqref{tr_0}}&\eeqref{eq1}
\end{diagram}}

Eq. \eeqref{mod_INB} with $a=0$ and \eeqref{mikh1} with $a=0$ are  simple modifications of the INB. The corresponding transformations are linearizable, and these equations are presented  in \cite{mx13} and \cite{b91}, respectively. Eq. \eeqref{mikh1} with $a=1$ has been found in \cite{ap14,mx13} together with corresponding transformations \teqref{minb}. As it is shown in \cite{ap14,gyl16}, these transformations are of Miura type.

Eq. \eeqref{eq1} is discussed in \cite{gyl16} as a preliminary result of the present classification. It is shown there that the transformation \teqref{tr_0} is linearizable, as it can be decomposed as a superposition of simple linearizable transformations, see Diagram \dref{d_compl}.

\eq{\begin{diagram} \label{d_compl}
\eqref{eq1_2}&\lTo{\eta_2: \teqref{tr_2}} &\eqref{eq1_1}\\
\dTo{\eta_1: \teqref{tr_1}}& &\dTo{\eta_1: \teqref{tr_1}}\\
\eeqref{INB}&\lTo{\eta: \teqref{tr_0}}&\eeqref{eq1}
\end{diagram}}
Eqs. (\ref{eq1_1},\ref{eq1_2})  used in this diagram have the form:
\begin{equation}\label{eq1_1} \dot u_0=(u_2-u_0)(u_1-u_{-1})(u_0-u_{-2}),\end{equation}
\begin{equation}\label{eq1_2} \dot u_0=u_0(u_1u_2-u_{-1}u_{-2}).\end{equation}
Eq. \eqref{eq1_2} is well-known \cite{b91}. It is interesting that this superposition consists of linearizable transformations in different directions, but it can be rewritten in the explicit form \teqref{tr_0}: 
$$\eta=\eta_1\circ\eta_2\circ\eta_1^{-1}.$$

\bigskip\bigskip

{ \centerline{{\bf List 5.} Other equations}}
\medskip
\begin{mequ}
{\dot u_0 = u_0^2(u_2u_1-u_{-1}u_{-2})-u_0(u_1-u_{-1})\label{seva}}
\end{mequ}
\begin{mequ}
{\dot u_0 = (u_0+1)\left(\frac{u_2u_0(u_{1}+1)^2}{u_1}-\frac{u_{-2}u_0(u_{-1}+1)^2}{u_{-1}}+(1+2u_0)(u_1-u_{-1})\right)\label{rat}}
\end{mequ}
\begin{mequ}
{\dot u_0 = (u_0^2+1)\left(u_2\sqrt{u_1^2+1}-u_{-2}\sqrt{u_{-1}^2+1}\right)\label{sroot}}
\end{mequ}
\bigskip
Equation \eeqref{seva} has been found in \cite{th96} and can be called  the discrete Sawada-Kotera equation \cite{th96,a11}. Equation \eeqref{sroot} has been found as a result of the present classification and seems to be new.
We know no relations between it and the other known equations. 
Eq. \eeqref{rat} has been found in \cite{a16_1} and is related to \eeqref{seva} as it is shown in the following diagram \cite{a16_1}:  
\eq{\begin{diagram} 
\eeqref{seva}&\lTo{\teqref{seva_to}}&\eqref{sh}&\rTo{\teqref{rat_to}}&\eeqref{rat}\label{diag_seva} 
\end{diagram}} Here
\eq{\label{sh}\dot u_0=\frac{(u_1-u_0)(u_0-u_{-1})(u_2-u_{-2})}{(u_2-u_{-1})(u_1-u_{-2})}.}

Introducing 
\eq{v_0=u_1-u_0,\qquad w_0=v_{0}/v_{-1},}
we can rewrite transformations \teqref{seva_to} and \teqref{rat_to} in form of the Riccati equations for unknown function $w_n$:
\eqs{\label{wuu}w_1w_0+\frac{1+\hat u_0}{\hat u_0}w_0+1=0,\quad w_1w_0+\frac1{1+\tilde u_1}w_1+w_0+1=0.}
This means, as shown in \cite{gyl16}, that, up to linearizable transformations, both transformations \teqref{seva_to} and \teqref{rat_to} are of Miura type. It is very difficult to use these relations \eqref{wuu} for the construction of solutions, as, if we know one of the solutions $\hat u_n$ or $\tilde u_n$ and need to find the other one, we have to solve a Riccati equation.

Let us explain now why \eeqref{seva}-\eeqref{sroot} cannot be transformed into the INB by an explicit non-point transformation.
Transformation $\tilde u_0=au_0,\ \tilde t =a^{-3}t$ introduces into equations \eeqref{seva}-\eeqref{sroot} the parameter $a$, e.g. in case of \eeqref{seva} it gives:
$$\dot u_0 = u_0^2(u_2u_1-u_{-1}u_{-2})-a^2u_0(u_1-u_{-1}).$$ All these three equations with the parameter $a$ are analogues of \eeqref{mikh1} as 
they generalize the same equation. More precisely, if we substitute $a=0$, then in all cases we obtain \eeqref{mikh1} with $a=0$. This is a well-known modification of the INB equation, see List 4. Equation \eeqref{mikh1} is transformed into the INB by \teqref{minb},  and it is natural to search analogoues transformations for  \eeqref{seva}-\eeqref{sroot}.

Let us consider a more general equation: 
\eq{\dot u_0=a(u_0)b(u_1)u_2-a(u_0)b(u_{-1})u_{-2}+c(u_1,u_0,u_{-1}\label{eq_gen}), }where the function $a(x)/b(x)$ is not constant. This equation generalizes \eeqref{mikh1}, \eeqref{seva}-\eeqref{sroot}. Let us look for a transformation of the form:
\eq{\hat u_0=\phi(u_k,u_{k-1},\ldots,u_0),\quad k>0,\quad \frac{\partial\phi}{\partial u_k}\neq0,\quad \frac{\partial\phi}{\partial u_0}\neq0,\label{zamg}}
which can be obtained with no loss of generality from \eqref{transf} by a shift. If we consider a transformation \eqref{zamg} which transforms  \eqref{eq_gen} into the INB \eeqref{INB}, then the function $\phi$ must satisfy the differential-functional equation
\eq{\dot \phi\equiv\sum_{j=0}^k\frac{\partial \phi}{\partial u_j}\dot u_j=\phi(T^2+T-T^{-1}-T^{-2})\phi,\label{eq_phi}}where $\dot u_j$ are defined by shifting $j$ times  \eqref{eq_gen}. 

 For all $k\neq 2$ it is sufficient to use two consequences of \eqref{eq_phi}, which are obtained by differentiation with respect to $u_{k+2}$ and $u_{-2}$:
$$\frac{\partial \phi}{\partial u_k}a(u_k)b(u_{k+1})=\phi T^2\frac{\partial \phi}{\partial u_k}, $$
$$\frac{\partial \phi}{\partial u_0}a(u_0)b(u_{-1})=\phi T^{-2}\frac{\partial \phi}{\partial u_0}.$$
By a short calculation it is easy to show that the  transformation \eqref{zamg} does not exist for all $k\neq2$. The case $k=2$ requires a more detailed investigation, as a solution of the problem exists in case of \eeqref{mikh1}. We are led to the following result:

\begin{theorem}\label{th_tran}
Eqs. \eeqref{seva}-\eeqref{sroot} cannot be transformed into the equation INB \eeqref{INB} by a transformation of the form \eqref{zamg}.
\end{theorem}
In conclusion, let us briefly discuss  the existence of a possible link between the various Lists 1-5. In List 5 we exclude from the consideration  \eeqref{sroot}, as at the moment we have no information about it. 

The $L-A$ pair for the Volterra equation \eeqref{Vol}, presented in List 1, is given by $2\times 2$ matrices as well as for  \eqref{our}, related to List 3, see \cite{gmy14}. The $L-A$ pair of the INB equation \eeqref{INB} and of the discrete Sawada-Kotera equation \eeqref{seva}, contained in List 4 and 5,  is given by $3 \times 3$ matrices, see \cite{bo88} and \cite{a11} respectively.
List 2 consists of linearizable equations. For this reason, three groups of equations, namely, List 2, Lists 1,3 and  Lists 4,5 should not be related by transformations \eqref{transf} and their compositions. 

Volterra type equations of List 1 essentially differ from relativistic Toda type equations of List 3 in their algebraic properties \cite{y06,asy00}. Then such equations cannot be related by the transformations mentioned above. As for equations of Lists 4 and  5, except for \eeqref{sroot}, at the moment we see no essential difference between them. We have a negative result for the explicit transformations \eqref{transf} formulated in Theorem \ref{th_tran}. However, there might be some compositions of transformations \eqref{transf}, analogues to the one shown in Diagram \eqref{diag_seva}, relating these Lists.

\section{  Generalized Symmetries of Key Equations}\label{5}


In this Section we present the generalized symmetries for the key equations of the  Lists 1-5. The symmetries for other equations can be easily obtained by using the simple transformations shown in Diagrams (\ref{dVol}, \ref{diag_lin}, \ref{diag_our}, \ref{dINB}) and contained in Appendix \ref{A4}. Those transformations either are contained in Table~\ref{t2} or are point equivalent to transformations of Table~\ref{t2}. 
The way how to construct the symmetries, using such transformations, is explained in Section \ref{s_con_l}. There is one exception which will be commented separately.

{\bf List 1.} Generalized symmetries for the Volterra equation and its modifications are well-known, see e.g. \cite{y06}, and we just replace $u_{n+j}$ by $u_{n+2j}$. \\
The simplest generalized symmetry for  \eeqref{Vol} reads:
\seq{u_{0,\tau}=u_0(u_2(u_4+u_2+u_0)-u_{-2}(u_0+u_{-2}+u_{-4})).}
The symmetry 
\seq{u_{0,\tau}=(u_0^2+cu_0)((u_2^2+cu_2)(u_4+u_0+c)-(u_{-2}^2+cu_{-2})(u_0+u_{-4}+c))}
corresponds to \eeqref{Vol0}, if $c=0$, and to \eeqref{Vol1}, if $c=1$.
Generalized symmetries of  \eeqref{Vol_mod}-\eeqref{Vol2} are constructed by using the transformations \teqref{tr_sum}-\teqref{tr_1}.

{\bf List 2.} A generalized symmetry for  \eeqref{Bur} could be obtained from the linear equation \eqref{lin_eq} by applying the transformation \teqref{tr_1}. However, we write it down here in explicit form:
\eq{u_{0,\tau}=(T^4-a^4)\left(\frac{u_{-1}u_{-3}}{u_{-2}}+\frac{u_0u_{-2}u_{-4}}{u_{-1}u_{-3}}\right).\label{s_Bur}}
A symmetry for \eeqref{Bur2} can be constructed by the transformation \teqref{tr_summ}. Eq. \eqref{s_Bur} can be represented as \eq{u_{0,\tau}=(T+a)h,\label{**}} and the symmetry for \eeqref{Bur2} is obtained from \eqref{**} as follows:
\seq{u_{0,\tau}=h|_{u_k\to u_{k+1}+au_k+b}.}

{\bf List 3.} One could get a generalized symmetry of \eeqref{our1}, using the known symmetry of \eqref{our} \cite{gy12,gmy14} and the transformation \teqref{tr_1}. However, we write it down explicitly:
\seq{u_{0,\tau}=(u_0-1)(T-T^{-1})\left((T+T^{-1})s_1-(1+T^{-1})s_2+s_3\right),}
\seq{s_1=\frac{u_{2}u_0u_{-2}(u_1-1)(u_0-1)(u_{-1}-1)}{u_1u_{-1}},\quad s_2=\frac{u_1u_{-1}(u_1-1)(u_0-1)}{u_0},}
\seq{s_3=\frac{u_1(u_0-1)(u_1u_0-u_1-u_0)u_{-1}^2}{u_0^2}-\frac{(u_2u_1-u_2-u_1)u_0(u_0-1)}{u_1}.}
A symmetry for  \eeqref{our2} can be obtained from this one by the transformation \teqref{tr_prod_m}.

{\bf List 4.} Generalized symmetries for almost all equations of this List are known;  for the INB equation \eeqref{INB} and for \eeqref{mikh1} with $a=0$ see \cite{ztof91}, and for   \eeqref{mikh1} with $a=1$ and for  \eeqref{mod_INB} with $a=0$  see \cite{mx13}. Nevertheless, for completeness, we present here the generalized symmetries for the most interesting equations of this List.
For  \eeqref{INB} the symmetry has the form
\seq{u_{0,\tau}=u_0\big(T^2+T-T^{-1}-T^{-2}\big)\big((T+T^{-1})u_1u_{-1}+(T+1)u_0u_{-1}+u_0^2\big),}
for  \eeqref{mikh1} the form
\seq{u_{0,\tau}=(u_0^2+au_0)\big(T^2-T^{-1}\big)\Big(u_0u_{-1}(1+T^{-1})\big((u_2u_1+u_{-1}u_{-2})(u_0+a)\big)+au_0^2u_{-1}^2\Big),} and for \eeqref{eq1} the form
\seqs{u_{0,\tau}=\frac{(u_1-u_0)(u_0-u_{-1})}{u_0}\left(\big(T^2-T^{-2}\big)\frac{u_2(u_1-u_0)(u_0-u_{-1})u_{-2}}{u_{1}u_{-1}}\right.\\\left.+\big(T-T^{-1}\big)\frac{u_1(u_1-u_0)(u_0-u_{-1})u_{-1}}{u_0^2}-p_1+p_2\right),} \seq{p_1=\frac{u_3(u_2-u_1)u_0u_{-2}}{u_2u_{-1}}+\frac{u_2u_0(u_{-1}-u_{-2})u_{-3}}{u_1u_{-2}},} \seq{p_2=\frac{(u_2u_{-1}-u_1u_{-2})(u_1u_0+u_0u_{-1}-u_1u_{-1})}{u_1u_{-1}}.} The symmetry for \eeqref{mod_INB} is constructed by using the transformation \teqref{tr_raz}.

{\bf List 5.} The generalized symmetry for the discrete Sawada-Kotera equation \eeqref{seva} has been found in \cite{a11} and it has the form:
\seqs{u_{0,\tau}=u_0\big(w_1(w_3+w_2+w_1+w_0)-w_{-1}(w_0+w_{-1}+w_{-2}+w_{-3})\\-u_{1}(w_{3}+w_{-1})+u_{-1}(w_1+w_{-3})\big),\quad w_0=u_0(1+u_1u_{-1}).}
For  \eeqref{rat} one has:
\seqs{u_{0,\tau}&=(u_0+1)\big((T^2-T^{-2})A+(T-T^{-2})B+(T-T^{-1})C+(1-T^{-1})D),\\
A&=U_{1}U_{-1}u_{2}u_{-2}u_0(1+u_0)+U_0u_{1}u_{-1},\\
B&=u_{2}u_{-1}((u_0U_1+1)(u_1U_0+1)+u_0^2U_1+u_1^2U_0-1)\\&+\frac{u_2u_0^2}{u_1}+\frac{u_1^2u_{-1}}{u_0}+u_1u_0,\\
C&=(U_0u_1u_{-1}+1)^2+U_0u_0(1+u_1u_{-1}(2+3u_1+3u_{-1}))\\&+u_1u_0u_{-1}(2(u_0+1)-3(u_1+u_{-1})),\\
D&=U_1U_0u_2u_1u_0u_{-1}+U_2u_3u_1u_0(u_1+1)+U_{-1}u_1u_0u_{-2}(u_0+1)+U_1u_2u_0^2\\&+U_0u_1^2u_{-1}+u_1^2(u_3u_2+2u_3+u_2+2u_{-1})+u_0^2(u_{-1}u_{-2}+u_{-1}+2u_{-2}+2u_2)\\&+u_1^2u_0(4u_{-1}-2u_2^2+3)+u_1u_0^2(4u_2-2u_{-1}^2+3)+2u_1^2u_0^2+5u_1u_0,\\
U_0&=\frac{(u_0+1)^2}{u_0}.}
For  \eeqref{sroot} the symmetry reads:
\seqs{u_{0,\tau}&=U_0\Big(u_{4}\sqrt{U_{3}}U_{2}\sqrt{U_1}-\sqrt{U_{-1}}U_{-2}\sqrt{U_{-3}}u_{n-4}\\&+u_2u_1\sqrt{U_1}\big(u_3\sqrt{U_2}+\sqrt{U_0}u_{-1}\big)-\sqrt{U_{-1}}u_{-1}u_{-2}\big(u_1\sqrt{U_0}+\sqrt{U_{-2}}u_{-3}\big)\\&+u_0\big(u_2^2U_1-U_{-1}u_{-2}^2\big)\Big),\qquad U_0=u_0^2+1.}

\section{Conclusion} \label{s_con}

In this article we have done the classification of the differential-difference equations depending on five lattice points and belonging to Class~I. This Class is a natural subclass of the differential-difference equations \eqref{e1} from the point of view of the integrability conditions. In this Class we have found 17 equations, some of which seem new.  We have found the non-point transformations which relate them to fewer key equations and presented for them the generalized symmetries.

This work is the starting point of a research which we plan to continue.  From one side, using the results presented here,  we can extend the classification outside Class~I by considering the case when \eqref{e2} is not satisfied. We can call this class of equations as a Class~II. We already have an example belonging to this class, namely \eqref{our}. From the other side, we can construct the B\"acklund transformations for differential-difference equations obtained in this paper. They possibly will provide autonomous integrable partial difference equations, defined on a square lattice, different from the known examples, see e.g. \cite{ABS,gy12,ly11,mx13,a11}.

\appendix
\appendixpage

\section{Some remarks on non-point transformations of differential-difference equations}\label{tran}
In this Appendix we follow the paper \cite{gyl16} and briefly formulate some results of the transformation theory necessary for this paper. In particular, we introduce the notions of linearizable and Miura type transformations used here. We also explain how to find transformations relating equations obtained. Schemes of the construction of modified equations by using those transformations are presented in Appendices \ref{s_point}, \ref{s_con_l} in a new way compared to \cite{gyl16}. Then we present the list of transformations necessary to reduce the list of equations to their key representatives.

\subsection{Miura type and linearizable transformations }\label{A1}

There are two essentially different classes of transformations \eqref{transf}. The first one consists of {\it Miura type transformations}. The inversion  of such transformations is equivalent to solving the discrete Riccati equations and their generalizations \cite{gyl16}.
For example the well-known discrete Miura transformation 
\eq{\hat u_0=(1+u_0)(1-u_{1})\label{dM}} relates the Volterra equation to its modification. Eq.  \eqref{dM} may be considered as a discrete Riccati equation for the unknown function $u_n$ with $\hat u_n$ a given function.

The second class consists of {\it linearizable transformations} which are not of Miura type.
We will use bellow {\it linear transformations} with {\it constant coefficients}:  
\eq{\hat u_0=\nu_k u_{k}+\nu_{k-1}u_{k-1}+\ldots+\nu_m u_{m}+\nu,\ \ k>m. \label{l_tran}}
 We can introduce the following definition:

\begin{defn} \label{d1} A transformation of the form \eqref{transf} is called {\it linearizable} if it can be represented as a superposition of linear transformations \eqref{l_tran} and point transformations $\hat u_0=\psi(u_0)$. A superposition of linearizable transformations in different directions will be also called a linearizable transformation. 
\end{defn}

For example, the transformations \eq{\hat u_0^{+}=u_{1}u_0,\qquad \hat u_0^{-}=u_{1}/u_{0},\qquad \tilde u_0=u_2u_{1}u_0,\label{pr_dev}} which are of the form \eqref{transf}, are linearizable since they can be expressed as $$\hat u_0^{\pm}=\left(\exp \circ (T\pm 1)\circ \log \right)\,u_0,\qquad \tilde  u_0=\left(\exp \circ (T^2+T+ 1)\circ \log \right)\,u_0.$$ 

Transformations can be in different directions. As an example we have  the composition  
$$ B\leftarrow D \leftarrow C\to A,$$ consisting of linearizable transformations, which can be rewritten as a transformation $A\to B$ of the form \eqref{transf}, see Diagram \dref{d_compl}. In Diagrams \dref{diag_lin} and \dref{diag_our} we have superpositions of the form $A\to B\leftarrow C$. In these cases, linearizable transformations relating $A$ and $C$ are implicit in both directions. 

The inversion of any linearizable transformation is reduced to solving a number of linear equations with constant coefficients. For this reason the linearizable transformation is not of Miura type and the construction of solutions by such transformations is more easy. 

Any linearizable transformation of the form \eqref{transf} can be expressed as a composition of the transformation \eq{\hat u_0=(T-1)u_0\label{ls}} and non-autonomous point transformations $\hat u_n=\psi_n(u_n)$ \cite{gyl16}. So, in order to invert a linearizable transformation, we have to solve a few times the simplest discrete linear equation \eqref{ls} for the unknown function $u_n$. The transformation \eqref{ls}  is solved by the discrete analogue of the integration.

Next, we will explain how to construct simple autonomous linearizable transformations, according to Definition \ref{d1},   \eq{\label{non_inver}\hat u_0=\phi(u_{1},u_0)} by using  {\it point symmetries} and {\it conservation laws}. These techniques will be hereinafter used to find links between equations of the resulting list presented in Section \ref{sec_list}.

\subsection{Point symmetries}\label{s_point}
Here we construct the transformations \eqref{non_inver}, starting from point symmetries.

 For an equation of the form \eqref{y0} we can describe all non-autonomous point symmetries of the form \eq{\label{psym}\partial_\tau u_n=\sigma_n(u_n), \quad \sigma_n(u_n)\neq0, \; \forall n,}  by solving the determining equation: 
 \eq{\sigma_n'(u_n)\Psi=\sum_{j=-2}^2 \frac{\partial \Psi}{\partial u_{n+j}}\sigma_{n+j}(u_{n+j}),\label{det_eq}} where by a $\sigma_n'$ we mean the derivative of the function with respect to its argument.

\begin{thm} \label{th1}
If  \eqref{y0} has a point symmetry \eqref{psym}, then it admits the following non-autonomous linearizable transformation:
 \eq{\hat u_n=(T-1)\eta_n(u_n),\quad \eta'_n(u_n)=\frac1{\sigma_n(u_n)},\label{trans_th}}
 which allows us to construct a modification of \eqref{y0}.
\end{thm}

We are interested here in autonomous linearizable transformations of the form \eqref{non_inver}, and therefore primarily in autonomous point symmetries \eqref{psym}. However, sometimes a non-autonomous point symmetry of the form \eqref{psym} may also lead to the autonomous result. This is the case when there exists a non-autonomous point transformation \eq{\label{p26}U_n=\psi_n(\hat u_n)} which turns the transformation \eqref{trans_th} into an autonomous one. We  guarantee that, in this case, the resulting equation for $U_n$ will be  also autonomous \cite{gyl16}. 

For any equation \eqref{y0} and any transformation \eqref{non_inver}, we can get an equation of the form:
\eq{\label{tty0}   \hat u_{0,t}=\hat \Psi(\hat u_{2},\hat u_{1}, \hat u_0, \hat u_{-1}, \hat u_{-2},u_0).} 
To do so we differentiate \eqref{non_inver} w.r.t. to $t$ in virtue of \eqref{y0} and express all variables $u_k,\ k\ne0,$ in terms of $\hat u_j,u_0$ by using \eqref{non_inver}. In the case of the autonomous transformations obtained with a help of Theorem \ref{th1} and the remark just after it, the dependence on $u_0$ in \eqref{tty0} disappears, and one gets in this way an autonomous modified equation \eqref{ty0}.

Let us write down in Table \ref{t1} the four most typical examples of point symmetries, two of which not autonomous, together with the corresponding autonomous transformations \eqref{non_inver}. Some of them have been simplified by applying an autonomous point transformation of $\hat u_0$.

\begin{table}[ht]
\begin{center}
\begin{minipage}{0.57\textwidth}
\caption{\small Examples of point symmetries \eqref{psym} and corresponding autonomous linearizable transformations \eqref{non_inver}  }
\label{t1}
\end{minipage}
		\begin{tabular}{|c|c|c|c|c|}
			\hline
			$\sigma_n(u_n)$ & $1$ & $(-1)^n$& $u_n$& $(-1)^nu_n$ \\
			\hline
			$\phi(u_{1},u_0)$ & $u_{1}-u_{0}$& $u_{1}+u_{0}$ &$u_{1}/u_{0}$&$u_{1}u_{0}$\\
			\hline
		\end{tabular}
\end{center}
\end{table}

\subsection{Conservation laws}\label{s_con_l}
We can construct simple autonomous linearizable transformations \eq{\label{non_inver1} u_0=\psi(\tilde u_{1}, \tilde u_0),}  starting from conservation laws. Transformation \eqref{non_inver1} relates equation \eqref{y0} and an equation of the form \eq{\label{hy0}   \tilde u_{0,t}=\tilde \Psi(\tilde u_{2},\tilde u_{1}, \tilde u_0, \tilde u_{-1}, \tilde u_{-2}).}
For any equation \eqref{y0} we can find all conservation laws of the form \eq{\label{claw}\partial_t\rho_n(u_n)=(T-1)h_n,\quad \rho_n'(u_n)\neq0, \; \forall n,}  where $h_n=h_n(u_{n+1},\ldots,u_{n-2})$. The conserved density $\rho_n$ is found  by using a criterion introduced in  \cite{ly97}. A function $\rho_n(u_n)$ is a conserved density  of  \eqref{y0} iff 
\eq{\frac{\delta(\partial_t\rho_n(u_n))}{\delta u_n}\equiv\sum_{j=-2}^2T^{-j}\frac{\partial (\rho_n'(u_n)\Psi)}{\partial u_{n+j}}=0.} If $\rho_n$ is known, then the function $h_n$ can be easily constructed \cite{ly97}.

\begin{thm} \label{th2}
If  \eqref{y0} has a conservation law \eqref{claw}, then it admits the following non-autonomous linearizable transformation:
 \eq{u_n=\rho_n^{-1}(\tilde u_{n+1}-\tilde u_n),\label{trans_th2}}
which allows us to construct a modification of \eqref{y0}:
\eq{\partial_t \tilde u_n=h_n(\rho_{n+1}^{-1}(\tilde u_{n+2}-\tilde u_{n+1}),\ldots,\rho_{n-2}^{-1}(\tilde u_{n-1}-\tilde u_{n-2}))\label{claw1}.}
\end{thm}

In the case of autonomous conservation law \eqref{claw}, we get an autonomous modified equation \eqref{hy0}.
When a non-autonomous point transformation \eq{\label{claw2} U_n=\mu_n(\tilde u_n)} makes the transformation \eqref{trans_th2} autonomous, we often can get an autonomous modification too, see \cite{gyl16}. 
In the particular case when  \eq{\rho_n(u_n)=(-1)^n p(u_n)\label{den_p},} the transformation \eqref{claw2} has the form
$U_n=(-1)^{n+1}\tilde u_n$, hence $$u_n=p^{-1}(U_{n+1}+U_n),$$ and we are guaranteed that there exists an autonomous modification \cite{gyl16}.

Let us write down in Table \ref{t2} the four most typical examples of conserved densities, two of which not autonomous and of the form \eqref{den_p}, together with corresponding autonomous transformations \eqref{non_inver1}. Some of them have been simplified by autonomous point transformations of $\tilde u_0$.
\begin{table}[ht]
\begin{center}
\begin{minipage}{0.64\textwidth}
\caption{\small Examples of conserved densities of \eqref{claw} and  corresponding autonomous linearizable transformations \eqref{non_inver1}}
\label{t2}
\end{minipage}
		\begin{tabular}{|c|c|c|c|c|}
			\hline
			$\rho_n(u_n)$ & $u_n$ & $(-1)^nu_n$& $\log u_n$& $(-1)^n\log u_n$ \\
			\hline
			$\psi(\tilde u_{1},\tilde u_0)$ & $\tilde u_{1}-\tilde u_{0}$& $\tilde u_{1}+\tilde u_{0}$ &$\tilde u_{1}/\tilde u_{0}$&$\tilde u_{1}\tilde u_{0}$\\
			\hline
		\end{tabular}
\end{center}
\end{table}

Let us explain how to find the modified equation for these examples. For the first and second ones we have to obtain the representations:
\seq{\dot u_0=(T\mp1)h,} and then we get the modification $\tilde u_{0,t}=h,$ where the variables $u_k$ in $h$ are replaced by $\tilde u_{k+1}\mp\tilde u_k$.
For the third and fourth examples we have to write down the representations
\seq{\partial_t\log u_0=(T\mp1)h.}  A modification has the form $\tilde u_{0,t}=\tilde u_0 h,$ where the variables $u_k$ in $h$ are replaced by $\tilde u_{k+1}\tilde u_k^{\mp1}$.

In conclusion, we note that if for a given conserved density $\rho_n(u_n)$ we get the representation \eqref{claw} with a different linear difference operator with constant coefficients,  then we can get a different linearizable transformation and corresponding modified equation in a similar way.
For instance, for the  conserved density $\log u_n$ of the INB equation \eqref{INB0} we have:
\seqs{\partial_t \log u_0=&(T^4+T^3-T-1)u_{-2}\\=&(T-1)(T+1)(T-c_1)(T-c_2)u_{-2},\nonumber} where $c_{1,2}=-\frac12\pm\frac{\sqrt{3}i}{2}$, and this provides a lot of possibilities to construct modified equations, see details and other examples in \cite{gyl16}.

\subsection{List of non-point transformations }\label{A4}

Here we list all non-point transformations used to link the equations presented in Section \ref{sec_list}:
\begin{transform}{\hat u_0=u_1+u_{0},\label{tr_sum}}\end{transform}
\begin{transform}{\hat u_0=u_1-u_0+a,\label{tr_raz}}\end{transform}
\begin{transform}{\hat u_0=u_1u_0 ,\label{tr_1}}\end{transform}
\begin{transform}{\hat u_0=u_2u_0, \label{tr_pr2}}\end{transform}
\begin{transform}{\hat u_0=u_2(u_0+1)\quad \hbox{ or }\quad \hat u_0=(u_2+1)u_0,\label{tr_m2}}\end{transform}
\begin{transform}{\hat u_0=u_1+au_{0}+b,\label{tr_summ}}\end{transform}
\begin{transform}{\hat u_0=u_1u_0+1 ,\label{tr_prod_m}}\end{transform}
\begin{transform}{\hat u_0=u_2u_1(u_0+a)\quad \hbox{or}\quad \hat u_0=(u_2+a)u_1u_0,\label{minb}}\end{transform}
\begin{transform}{\hat u_0=\frac{(u_1-u_0)(u_0-u_{-1})}{u_0},\label{tr_0}}\end{transform}
\begin{transform}{\hat u_0=u_1-u_{-1}\label{tr_2},}\end{transform}
\begin{transform}{\hat u_0=\frac{u_0-u_1}{u_2-u_{-1}},\label{seva_to}}\end{transform}
\begin{transform}{\tilde u_0=-\frac{(u_1-u_{-1})(u_0-u_{-2})}{(u_1-u_{-2})(u_0-u_{-1})}.\label{rat_to}}\end{transform}

\medskip
\paragraph{Acknowledgments.}  The authors RNG and RIY gratefully acknowledge financial support from a Russian Science Foundation
grant (project 15-11-20007). DL has been partly supported by the Italian Ministry of Education
and Research, 2010 PRIN {\it Continuous and discrete nonlinear integrable evolutions: from
water waves to symplectic maps} and by INFN IS-CSN4 {\it Mathematical Methods of Nonlinear
Physics}.

\end{document}